\documentclass[aps,preprint,showpacs,epsfig]{revtex4}
\usepackage{amsmath}
\usepackage{epsfig}

\begin{document}
\bibliographystyle{/usr/share/texmf/tex/latex/revtex4/prsty}

\title{
Effect of P-wave interaction in $^6$He and $^6$Li photoabsorption}

\author{Sonia Bacca$^{1,2}$, Nir Barnea$^3$, Winfried Leidemann$^{1}$ and Giuseppina Orlandini$^{1}$}

\affiliation{$^1$
Dipartimento di Fisica, Universit\`a di Trento and INFN (Gruppo Collegato
di Trento),\\via Sommarive 14, I-38050 Povo, Italy} 
\affiliation{$^2$ Institut f\"ur Kernphysik, Johannes-Gutenberg Universit\"at
Mainz,\\ Johann-Joachim-Becher Weg 45, D-55099 Mainz, Germany}
\affiliation{$^3$Racah Institute of Physics, Hebrew University, 91904, 
Jerusalem, Israel}

\date{\today}

\begin{abstract}
The total photoabsorption  cross sections of the six-body nuclei are 
calculated including complete final state interaction via the Lorentz Integral 
Transform method. The effect of nucleon-nucleon central P-wave forces is 
investigated. Comparing to results with central potentials 
containing S-wave forces only, one finds considerably more strength in the 
low-energy cross sections and a rather strong improvement in comparison with 
experimental data, in particular for $^6$Li.
\end{abstract}
\bigskip
\maketitle

\vfill\eject
In a recent paper \cite{Sonia02} we carried out the first microscopic 
calculation of total photoabsorption cross sections for $A=6$ nuclei.
Semi-realistic central S-wave forces were taken as NN interaction. The 
obtained cross sections were not in good agreement with the low-energy data. 
Since $^6$He and $^6$Li are P-shell nuclei we suspected in \cite{Sonia02} that 
this observable could show a sensitivity to the P-wave NN interaction. The aim 
of the present work  is to check this conjecture. To this end we use the 
recently published AV4' potential \cite{AV4'}, which includes S- and P-waves 
forces. 

Our calculation proceeds in the same way as in \cite{Sonia02}. Here we 
summarize the main steps. The total photoabsorption cross section in unretarded
dipole approximation is given by 
\begin{equation}
\sigma(\omega)=4\pi^{2}\alpha\omega R(\omega)\,,
\end{equation}
\noindent where $\alpha$ is the fine structure constant, $\omega$  the 
photon energy, and 
\begin{equation}
{\label{1}
R(\omega )=\int d\Psi _{f}\left| \left\langle \Psi _{f}\right| \hat{D}_z \left| 
\Psi _{0}\right\rangle \right| ^{2}\delta (E_{f}-E_{0}-\omega)}
\end{equation}
\noindent  the response function; wave functions and energies of ground and final state are denoted by  $\left| \Psi_{0/f} \right>$ and 
$E_{0/f}$ respectively, while  
\begin{equation}
\label{dip}
\hat{D}_{z}=\sum_{i=1}^{A}\frac{\tau^{3}_{i}z'_{i}}{2}\,
\end{equation}
is the unretarded dipole operator. Here $\tau^{3}_{i}$  and $z'_{i}$ 
represent the third component of the isospin operator and of the spatial
coordinate of the $i$-th particle in the center of mass frame, respectively. 
In the Lorentz Integral Transform (LIT) method 
\cite{LIT} one obtains $R(\omega)$ from the inversion of an integral transform
with a Lorentzian kernel
\begin{equation}
L(\sigma_{R},\sigma_{I} )= \int d\omega \frac{R(\omega )}
{(\omega -\sigma _{R})^{2}+\sigma ^{2}_{I}}= \left\langle 
\widetilde{\Psi }|\widetilde{\Psi }\right\rangle \,, \label{2}
\end{equation}

\noindent where the "Lorentz state" $\widetilde\Psi$ is the  
unique solution of an inhomogeneous ``Schr{\"o}dinger-like'' equation 
\begin{equation}
\label{3}
(H-E_{0}-\sigma_{R}+i\sigma_{I})|\widetilde{\Psi}\rangle=\hat{D}_{z}|
{\Psi_{0}}\rangle
\end{equation}
\noindent 
with asymptotic boundary conditions similar to a bound state. Thus one can apply
bound-state techniques for its solution. We expand $\Psi_{0}$ and $\widetilde
{\Psi}$ in terms of the six-body symmetrized hyperspherical harmonics (HH) 
\cite{BN97}. The expansion is performed up to maximal values of the HH grand-angular momentum quantum number $K^0_{max}$ for $\Psi_0$ and   
$K_{max}$  for $\widetilde\Psi$. We improve the convergence 
of the HH expansion using the effective interaction hyperspherical harmonics 
(EIHH) approach \cite{EIHH}, where the bare potential is replaced by an 
effective potential constructed via the Lee-Suzuki method \cite{LeeS}. When 
convergence is reached, however, the same results are obtained as with the bare
potential (see Ref. \cite{EIHH}).

As in \cite{Sonia02}  we evaluate the LIT calculating the quantity $\langle 
\widetilde{\Psi }|\widetilde{\Psi}\rangle $ directly via the Lanczos 
algorithm \cite{Mar02}. We study the convergence of the LIT as a function of 
$K$. Our procedure consists in increasing  $K^0_{max}$ until convergence of the 
ground state is reached and then studying the behavior of 
the LIT with growing $K$. In case of the AV4' potential a sufficiently 
convergent result for the bound state is reached with $K^0_{max}= 12$ (yielding 
binding energies $E_0=32.90$ MeV for $^6$He and $E_0=36.47$ MeV for $^6$Li ). 
Since $\widetilde{\Psi}$ depends on $\Psi_0$ we also check 
whether the norm $\langle \widetilde{\Psi }|\widetilde{\Psi}\rangle $, i.e.
$L(\sigma_R,\sigma_I)$, converges for $K^0_{max}=12$. Indeed the transforms 
$L(\sigma_R,\sigma_I=10$ MeV)  obtained with $K^0_{max}=12$ and 14 at fixed 
$K$ differ by less than $1\%$. In case of a central S-wave interactions only, 
as for MTI-III \cite{MT69} and MN \cite{MN77} potentials, convergence is 
already reached at $K^0_{max}=10$.

\begin{figure}[ht]
\includegraphics{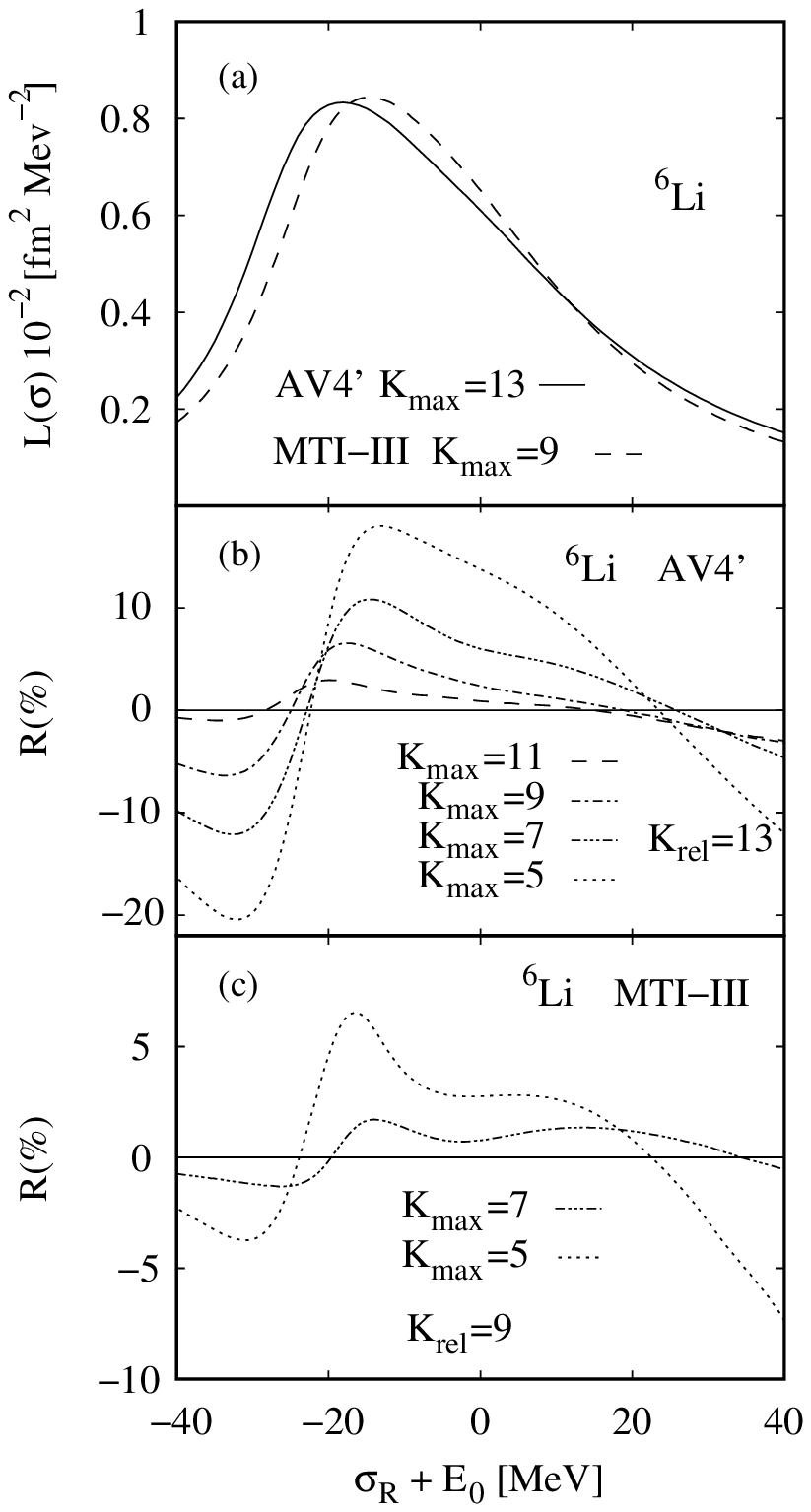}
\caption{(a) LIT for $^{6}$Li $(\sigma_I=10$ MeV) with AV4' and MTI-III 
potentials; HH convergence of LIT as function of $K$ with
 $K_{max}=13$ (see (6)) for AV4' (b) and MTI-III (c) potentials.}
\label{fig1}
\end{figure}

In Fig.~\ref{fig1} we show the convergence of the LIT for $^6$Li for the AV4' and the MTI-III
potentials. In the upper panel the two LIT results obtained with the highest 
considered $K_{max}$ are presented, while in the two lower panels we show the 
relative error $R$ in percentage, for the two potentials separately. The 
quantity $R(\%)$ is defined  as
\begin{equation}
R(\%)= \frac{(L(K)-L(K_{max}))}{L(K_{max})}\times 100 \,.
\end{equation}
One  can clearly see a rather nice convergence pattern with increasing $K$. 

\begin{figure}[ht]
\includegraphics{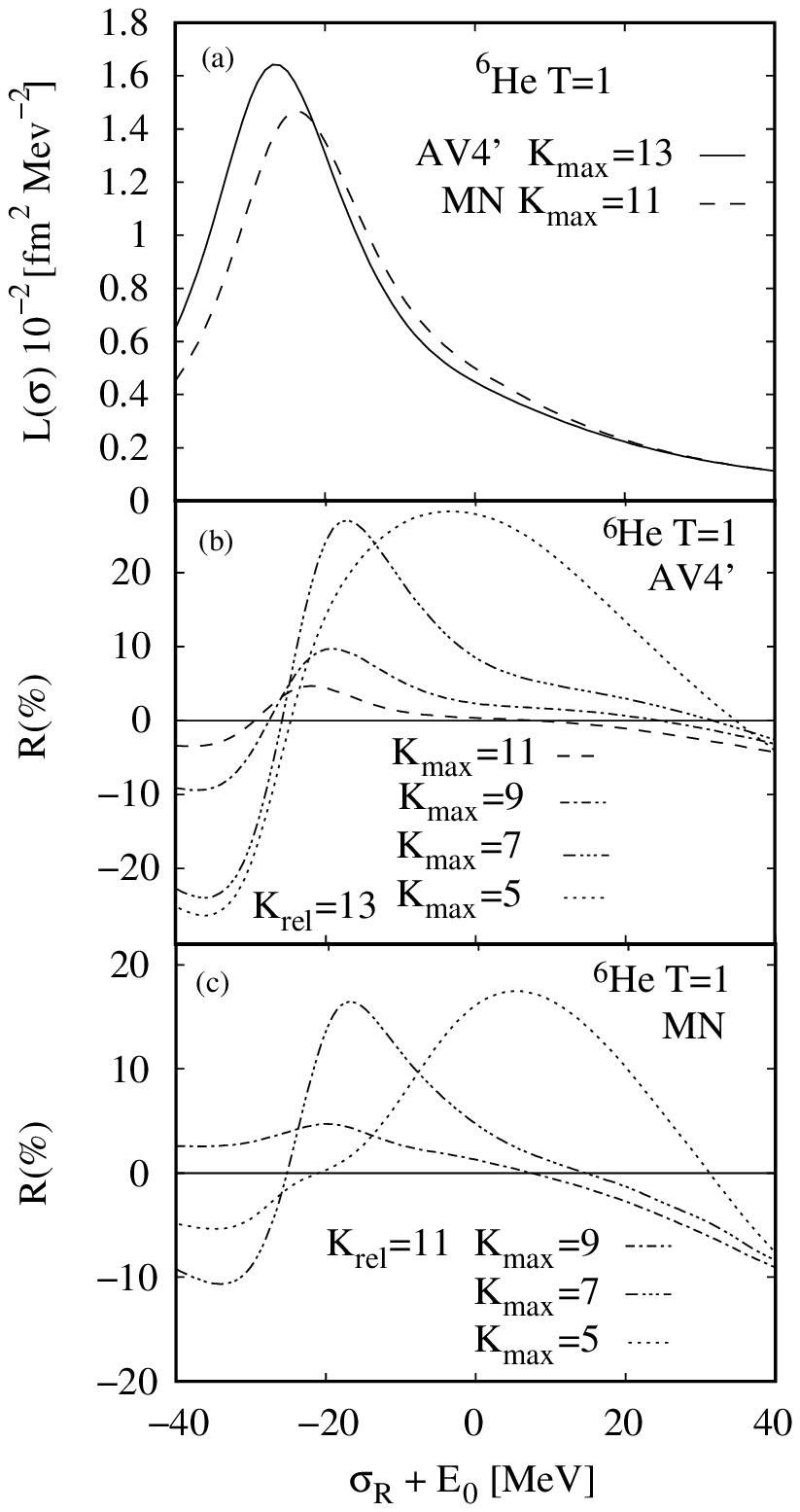}
\caption{(a) LIT for $^{6}$He $(\sigma_I=10$ MeV) with AV4' and MN 
potentials; HH convergence of LIT as function of $K$ with $K_{max}$ (see (6))
for AV4' (b) and MTI-III (c) potentials.}
\label{fig2}
\end{figure}

In Fig.~\ref{fig2} we present an analogous picture for $^6$He with the AV4' and the MN 
potentials. One finds rather satisfactory results, but compared to $^6$Li the 
AV4' case exhibits a slower convergence, e.g., in the lower $\sigma_R$ range,
where mainly strength from the threshold region is sampled, one has $R(K_{max}=
11) \simeq 1\%$ in case of $^6$Li and $R(K_{max}=11) \simeq 3\%$ in case of 
$^6$He. Figures 1 and 2 also illustrate 
that the convergence is better for pure S-wave potentials. Thus an 
addition of P-wave interaction seems to lead to a slightly weaker convergence
of the HH expansion. In fact performing LIT calculations with a modified 
AV4' potential, namely with switched off P-wave interaction, we get a 
convergence pattern similar to  those of MN or MTI-III potentials.

In the following we want to describe in some more detail the LIT calculation 
with $K_{max}=13$.
In Table I  we list the number $N$ of $^6$Li-HH basis states as a function of 
$K_{max}$. For the total number of basis functions one has to multiply $N$ by 
the number $N_{\rho}$ of hyperradial states ($N_{\rho}\approx 30$). Thus the 
total number of states becomes quite high and it is desirable to discard HH 
states which give only  negligible  contributions to the LIT. To this end we 
study the importance of the HH states according to their spatial symmetries. 
We find that quite a few  of them can be safely neglected beyond given 
$K_{max}$ values (see Table II). In this way for $K_{max}=13$ we accomplish  a 
reduction from $N=18402$ to $N=6362$. As one can see in Table II, the 
symmetries labelled [111111],[21111] and 
[3111] are not included at all  and others, namely [2211] and [321], are 
considered only up to $K^{sym}_{max}=7$ and $K^{sym}_{max}=11$, respectively.
We have checked the quality of our approximation, performing full 
calculations without cuts for lower $K$ values and comparing the obtained LIT 
results with those of a truncated calculation. When differences are negligible
we conclude that omitted states have no influence also in calculations with 
higher $K$. The omission of the K=13 states of symmetry [321] could not be 
checked in such a way, but already its K=11 contribution is almost negligible 
and thus its K=13 contribution should be of no importance. In an analogous 
way we carry out the calculation of the LIT for $^6$He. We would like to 
mention that for this nucleus one has two separate HH expansions for 
$\widetilde\Psi$, namely for the two isospin channels with T=1 and 
2~(see also \cite{Sonia02}).

After having discussed the convergence of the LIT we turn to the 
photodisintegration. In order to obtain the total photoabsorption cross section 
$\sigma(\omega)$ one has to invert the LIT of (4) (for details see 
\cite{ELO99}). This leads to the response function  $R(\omega)$ and thus to 
$\sigma(\omega)$, Eq. (1).
  
\begin{figure}[ht]
\includegraphics{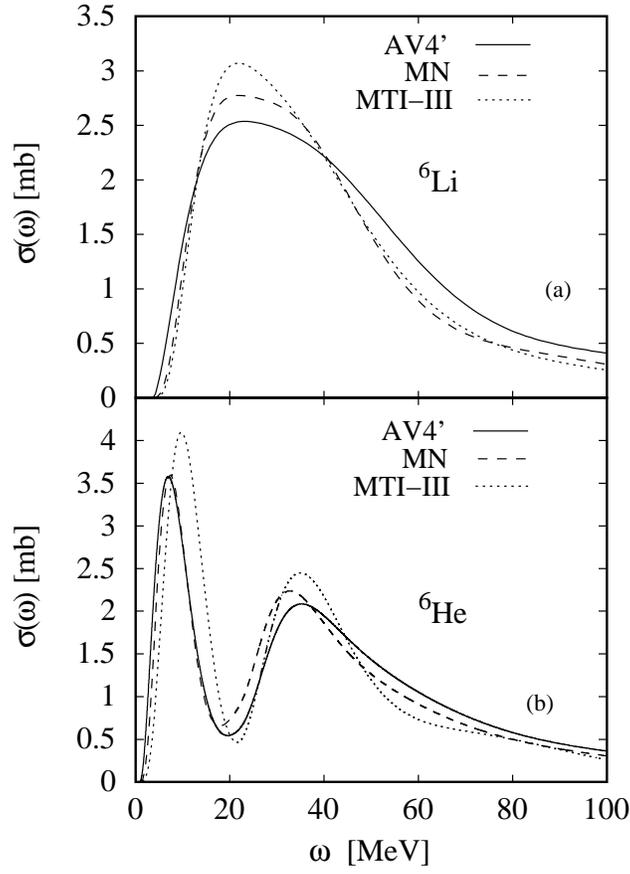}
\caption{Total photoabsorption cross sections for the six-body nuclei with  
AV4',  MN and MTI-III potentials: $^{6}$Li (a), $^{6}$He (b).}
\label{fig3}
\end{figure}

In Fig.~\ref{fig3} we show the results for the total photoabsorption cross 
section of $^6$Li and  $^6$He with the AV4' potential. In comparison we also 
present our previous results from \cite{Sonia02} with  MN and MTI-III 
potentials. One notes that the general structure of the cross section is 
similar for the various potential models, in particular the presence of two 
peaks for $^6$He, but one also finds 
potential dependent results for peak positions and peak heights. The double 
peak structure of $^6$He can be interpreted as a response of a {\it halo} 
nucleus, where the low-energy peak is due to the {\it halo}--$\alpha$ core
oscillation (soft dipole response) and the peak at higher energies due to the 
neutron-proton spheres oscillation (Gamow-Teller mode or hard dipole response).

\begin{figure}[ht]
\includegraphics{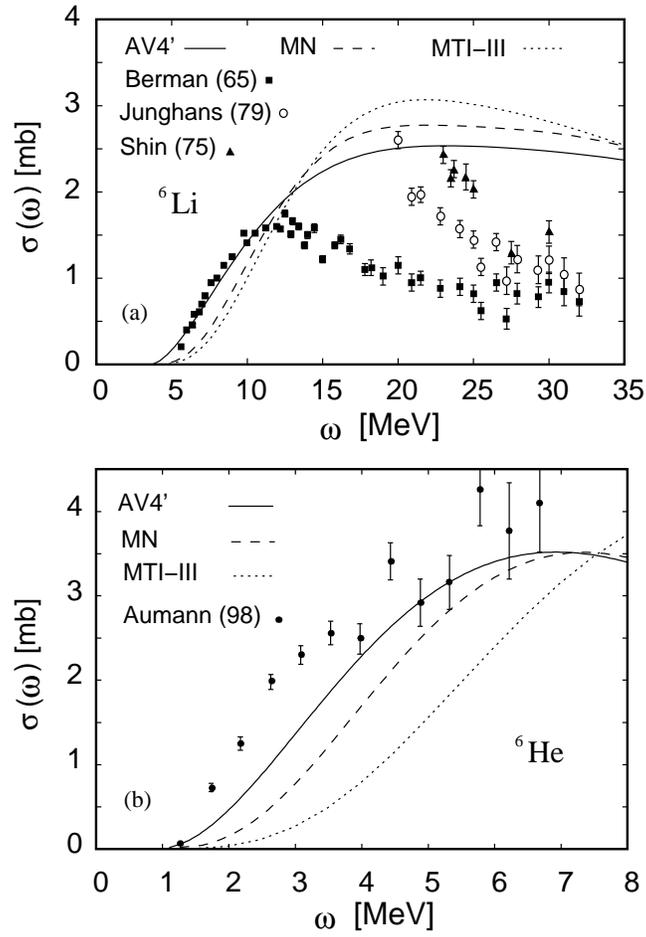}
\caption{Theoretical and experimental photoabsorption cross section results
(see also text): (a) $^{6}$Li with experimental data from \cite{Ber65, Jun79, 
Sh75}, (b) $^{6}$He with data from \cite{Au98, Aupriv} (theoretical results 
convoluted with instrumental response function).}
\label{fig4}
\end{figure}

In Fig.~\ref{fig4} we show the theoretical results together with available 
experimental data. 
Here we would like to mention that the data of \cite{Ber65} have been taken via
a semi-inclusive ($\gamma,n)$ measurement. The obtained results correspond
to the inclusive cross section up to an energy of about $15.7$ MeV, where 
additional channels open up. The cross section due to those additional channels 
have been measured in further experiments \cite{Sh75,Jun79}. In order to have 
an estimate for the total cross section we have simply summed the $(\gamma,n)$ 
data of \cite{Ber65} to the cross sections of \cite{Sh75,Jun79}. The data of 
Fig.~\ref{fig4} cited as \cite{Sh75,Jun79} are these sums.

Figure~\ref{fig4} shows that for the AV4' potential one finds an enhancement of strength 
in the threshold region compared to the S-wave potentials. It is evident
that the inclusion of the P-wave interaction improves the agreement with 
experimental data considerably. This is particularly the case for $^6$Li. In 
fact with the AV4' potential one has a rather good agreement with experimental 
data up to about 12 MeV. In case of $^6$He the increase of low-energy strength 
is not sufficient, there is still some discrepancy with data. Probably, in 
order to describe the {\it halo} structure of this nucleus in more detail 
additional potential parts are needed. In particular the spin-orbit component 
of the NN potential could play a role in the determination of the soft dipole 
resonance. In fact in a single particle picture of $^6$He the two {\it halo} 
neutrons will mainly stay in a p-state and can interact with one of the
core nucleons via the NN LS-force. Another reason for the discrepancy
could be the convergence. As already 
pointed out, our HH convergence is quite satisfactory for $^6$Li, whereas it 
is still not yet fully complete in case of $^6$He. The pronounced {\it halo}
structure of this nucleus could make the HH expansion more difficult.  
Nevertheless we would like to emphasize again that for all three potential 
models a typical $^6$He {\it halo} response appears automatically from a 
microscopic six-body calculation, while other details of the response are 
very sensitive to the interaction model.

In conclusion we can say that the P-wave interaction has an important
impact on the low-energy total photoabsorption cross section of the six-body
nuclei. It enhances the low-energy strength quite significantly. It also leads
to a considerable improvement in the comparison of theoretical and experimental
results, even if a fully satisfactory agreement is not yet reached. Further
investigations, both in theory and experiment, are needed.
As already pointed out in \cite{Sonia02} experimental data are too few ($^6$He)
or do not present a clear picture ($^6$Li). On the other hand,  from the 
theoretical point of view, more effort has to be addressed to the inclusion of 
additional parts in the NN potentials. Such future studies should lead to
a better understanding which NN potential parts are relevant in the 
six-nucleon photodisintegration.

\section*{Acknowledgements}
One of us (S.B.) thanks H. Arenh\"{o}vel for helpful discussions. The work of 
N.B. was supported by the Israel Science Foundation (grant no. 202/02).
A great parts of the numerical calculations were performed at the computer 
centre CINECA (Bologna). \\

\begin{table}
\caption{Number $N$ of $^6$Li HH basis states for various $K_{max}$ values.}
{\par\centering
\begin{tabular}{c|ccccc}
\hline
\hline
{$K_{max}$}&{5}&{7}&{9}&{11}&{13}\\
\hline
{$N$}&{~52~}&{~323~}&{~1489~}&{~5665~}&{~18402~}\\
\hline
\hline
\end{tabular}      
\par}
\end{table}

\begin{table}
\caption{Cut of symmetries for the $^6$Li calculation with $K_{max}=13$. 
For a given symmetry $N^{sym}$ denotes the number of included basis states 
and $K^{sym}_{max}$ is the maximal considered value of the grand-angular 
momentum quantum number for this symmetry.}
{\par\centering
\begin{tabular}{c|ccccccc}
\hline
\hline
{Symmetry}&{~[111111]~}&{~[21111]~}&{~[2211]~}&{~[3111]~}&{~[321]~}&{~[411]~}&{~[33]~}\\
\hline
{$N^{sym}$}&{0}&{0}&{50}&{0}&{2382}&{2598}&{1332}\\
\hline
{$K^{sym}_{max}$}&{--}&{--}&{7}&{--}&{11}&{13}&{13}\\
\hline
\hline
\end{tabular}      
\par}
\end{table}

\end{document}